# Neural Network Models for Software Development Effort Estimation: A Comparative Study

Ali Bou Nassif[1,a], Mohammad Azzeh[2,b], Luiz Fernando Capretz[3,c] and Danny Ho[4,d]

[1]Department of Electrical and Computer Engineering, University of Sharjah, Sharjah, UAE

[2]Department of Software Engineering, Applied Science University, Amman, Jordan, POBOX 166

[3]Department of Electrical and Computer Engineering, University of Western Ontario, London, ON, Canada

[4]NFA Estimation Inc., Richmond Hill, Ontario, Canada

[a]anassif@sharjah.ac.ae, [b]m.y.azzeh@asu.edu.jo, [c]lcapretz@uwo.ca, [d]danny@nfa-estimation.com

**Abstract:** Software development effort estimation (SDEE) is one of the main tasks in software project management. It is crucial for a project manager to efficiently predict the effort or cost of a software project in a bidding process, since overestimation will lead to bidding loss and underestimation will cause the company to lose money. Several SDEE models exist; machine learning models, especially neural network models, are among the most prominent in the field. In this study, four different neural network models – Multilayer Perceptron, General Regression Neural Network, Radial Basis Function Neural Network, and Cascade Correlation Neural Network – are compared with each other based on: (1) predictive accuracy centered on the Mean Absolute Error criterion, (2) whether such a model tends to overestimate or underestimate, and (3) how each model classifies the importance of its inputs. Industrial datasets from the International Software Benchmarking Standards Group (ISBSG) are used to train and validate the four models. The main ISBSG dataset was filtered and then divided into five datasets based on the productivity value of each project. Results show that the four models tend to overestimate in 80% of the datasets, and the significance of the model inputs varies based on the selected model. Furthermore, the Cascade Correlation Neural Network outperforms the other three models in the majority of the datasets constructed on the Mean Absolute Residual criterion.

## 1. Introduction

Estimating the cost or the effort[1] to develop a software application is crucial in the early stages of the software development life cycle. Many practitioners or project managers are notorious for underestimating the actual cost of the software. According to the International Society of Parametric Analysis (ISPA) [1] and the Standish Group International [2], two thirds of software projects fail to be delivered on time and within budget. The main two reasons behind software project failures are: (1) improper estimation in terms of project size, cost, and staff needed and (2) uncertainty of software and system requirements.

In the past three decades, different software cost estimation models have been developed [3]. The techniques that software cost predictors[2] use, fall into three main groups [4]. These groups are:

1. Expert Judgment: In this group, the software project predictor uses his or her expertise to predict the effort of such projects [5]. The expertise of the estimator is based on the problem domain, as well as on his or her familiarity with similar and historical projects.

2. Algorithmic (a.k.a. Parametric) Models: These models include COCOMO [6], SEER-SEM [7], and SLIM [8]. This group was very popular in the last three decades. Software size is the main input of these models

---

[1] The terms software cost and software effort are used interchangeably in this study

[2] The terms software cost estimation and software cost prediction are used interchangeably in this study

and the unit of software size is usually Function Points (FP) or Source Lines Of Code (SLOC). Linear and non-linear regression equations can be used in the algorithmic models.

3. Machine Learning: Machine learning has gained popularity in cost estimation predictive models. These models can be standalone models or models that work in conjunction with algorithmic models. Examples of standalone models include [9] in fuzzy logic, [10] [11] in neural networks. Examples of non-standalone models include [12] [13] [14] in fuzzy logic and [15] [16] in neuro-fuzzy.

There are some advantages of using non-parametric models over parametric models. These include:

• The relationship between the output and the input in parametric models is sometimes linear or nonlinear as the case in COCOMO. Although nonlinear models are more realistic in the world of software effort estimation, these nonlinearities are restricted to one or two equations in parametric models such as the case in Basic COCOMO and Intermediate COCOMO. However, non-parametric models, are linear, as well as, nonlinear models, that can map input to output without any restrictions.

• Parametric models are more sensitive to the nature of data distribution such as Gaussian distribution.

• In parametric models, the number of parameters is fixed. However, in non-parametric models, the number of parameters grow with training data.

• In order to use a specific parametric model (e.g. COCOMO), the model inputs (parameters and cost drivers of a project) must be known in advance. However, non-parametric models, such as Artificial Neural Network (ANN) models, can handle different number and types of inputs.

There are different algorithms available to train ANN models. The most popular one is the backpropagation algorithm that uses gradient descent in supervised learning. The algorithm calculates the gradient of a loss function with respect to all the weights in the network. The gradient is fed to the optimization method which in turn uses it to update the weights, in an attempt to minimize the loss function. Another training algorithm is the conjugate gradient algorithm [27] which uses the gradient during the backward propagation of errors through the network.

There is no unanimous agreement among researchers that any of the above groups works perfectly in all situations [17]. However, as mentioned above, machine learning models, especially neural network models, have gained a lot of popularity in this field [18]. As discussed in the Background and Related Work Section (Section 2), several types of neural network models have been used to predict the effort of software projects. These types are Multilayer Perceptron (MLP), General Regression Neural Network (GRNN), Radial Basis Function Neural Network (RBFNN), and Cascade Correlation Neural Network (CCNN). The main limitation of implementing these models is that there is no consensus on which model is the best. For instance, Nassif et al. [10] found that the MLP model excels when small-sized projects are being used. C. Lopez-Martin [19] found that the GRNN model surpasses multiple linear regression and fuzzy logic models. In other studies, C. Lopez-Martin [20] showed that the RBFNN model outperforms the MLP and the GRNN models. The CCNN was only applied in one study regarding software development effort estimation and it was found to outperform the multiple linear regression model [21].

There are three main reasons behind the inconsistency in the performance of the above models. (1) The datasets used in the model evaluation were different in each model. This is a main issue because the performance of such a model might be superior based on one dataset but it could be inferior based on other datasets. (2) Another main issue is the performance criteria used in the evaluation process. For instance, the predictive accuracy of a model can be remarkable on one criterion such as the Mean Magnitude of Relative Error (MMRE) but could be detrimental based on another criterion such as the Mean Absolute Residual (MAR). (3) The experimental and methodology setup might be different.

To address the above limitations, we carried out this research to compare four different neural network models named MLP, GRNN, RBFNN, and CCNN for a software development effort estimation using the same datasets and the same performance accuracy criterion. The datasets used in this study are industrial datasets from the International Software Benchmarking Standards Group (ISBSG) [22]. The performance accuracy criterion used is the MAR since other criteria such as MMRE are considered biased and not recommended in the evaluation of software cost estimation models [23]. The description of the datasets and the performance evaluation criterion will be discussed in Sections 3 and 4, respectively. In particular, we raise the following research questions:

RQ1: Which of the above neural network models (MLP, GRNN, RBFNN, and CRNN) has the lowest MAR?

In software estimation models, the prediction accuracy is inversely proportional to the MAR. This indicates that models with the highest accuracy have the lowest MAR.

RQ2: Which neural network model tends to overestimate and which tends to underestimate?

Overestimation and underestimation are side effects of improper estimation. We will test if a model overestimates or underestimates based on the comparison between the actual and the estimated efforts. Higher estimated efforts indicate that a model is overestimating. Likewise, lower estimated efforts is an indication of underestimation.

RQ3: Does the significance of model inputs vary from model to model?

Although software size is the most significant input in software cost estimation models, and this was used as a sole input in some models, M. Kassab [24] [25] showed that other inputs, such as non-functional requirements, can increase the software development effort by 100%. We will check if the importance of such input is different in different models.

This paper is structured as follows: Section 2 provides an overview of background and related work. Section 3 explains the datasets used in this study. Section 4 presents the performance evaluation criteria. Section 5 discusses the models training and testing, and gives the results of the comparison of neural network models. In Section 6, we discuss the results and answer the research questions. Section 7 lists some threats to validity, and lastly Section 8 concludes the paper and suggests future work.

## 2. Background and Related Work

Wen et al. [18] conducted a systematic literature review on machine learning based on SDEE between 1991 and 2010 and concluded that machine learning models are promising in the field of SDEE. Furthermore, the review showed that artificial neural network models are ranked second based on the number of selected publications.

An Artificial Neural Network (ANN) is a network that is comprised of nodes or artificial neurons that imitate the biological neurons [26]. ANN networks map an input to an output using linear or non-linear functions based on the nature of the problem and can be used as classifiers or regression. In SDEE, ANN models are used to solve a regression problem. Feed-forward networks are very popular in ANN where the flow of information is always from input to output. An ANN is usually composed of an input layer, hidden layer, and output layer. When an ANN model is used to predict the effort of a software project, the input layer is composed of nodes such as software size, programming language, programmer experience, etc. The middle layer contains a certain number of neurons, and each has a weight. The output layer contains one node that is the predicted effort. The output of a node is defined based on an activation function. This is represented mathematically in Equation (1). In the next subsections we will discuss the four types of neural networks that are used in our study.

$$y(t) = f[\sum_{i=1}^{n} w_i x_i - w_0]. \tag{1}$$

where $x_i$ are neuron inputs, $w_i$ are the weights and $f[.]$ is the activation function.

### 2.1 Multilayer Perceptron (MLP)

An MLP is a feed-forward typed artificial neural network model that has one input layer, at least one hidden layer, and one output layer. Each neuron of the input layer represents an input vector. If a network is only composed of an input layer and an output layer (no hidden layer), then the name of the network becomes Perceptron. In general, for an MLP network, a non-linear activation function is used in the neurons of the hidden layer. On the contrast, a linear activation function is usually used in the output layer. The number of the neurons in the hidden layer varies based on the number of input neurons and the type of the training algorithm used. One of the popular training algorithms is the backpropagation algorithm which is a type of gradient decent algorithm. Another algorithm that can be used to train a MLP network is the conjugate gradient algorithm [27]. In this study, we used the conjugate gradient algorithm because of the advantages it provides over the backpropagation algorithm [27]. Figure 1 shows the diagram of a MLP network that has five input vectors, seven neurons and one output.

MLP networks are among the most used neural network models in SDEE [28]. These MLP models are evaluated by their comparison with multiple linear regression models based on performance evaluation criteria such as MMRE.

## 2.2 Radial Basis Function Neural Network (RBFNN)

A RBFNN was first introduced in 1988 by Broomhead and Lowe [29]. An RBFNN network is a feed-forward network composed of three layers: an input layer, a hidden layer with a non-linear RBF activation function, and a linear output layer. Figure 2 shows the diagram of the RBFNN network.

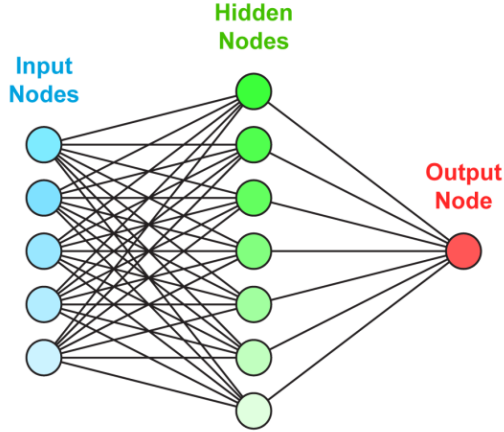
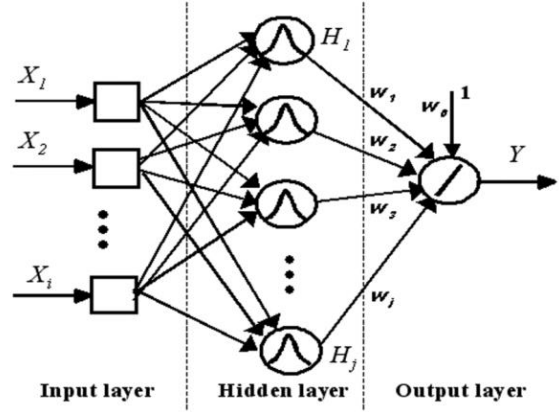

Figure 1: Multilayer Perceptron Model     Figure 2: Radial Basis Function Neural Network [5]

An RBFNN network mainly uses the Gaussian function and it depends on the distance from its center $C_i$ to the input $X$. Each RBF function has a radius called spread and is denoted by "$\sigma$". The spread might be different for each neuron. The mathematical representation of the RBF function is shown in Equation (2):

$$f(x) = \exp(-\frac{\|X - C_i\|^2}{2\sigma_i^2}).  \qquad (2)$$

Where $C_i$ is the center and $\sigma_i$ is the spread of the $i^{th}$ neuron in the hidden layer. The distance between the centre and $X$ is usually an Euclidean distance. RBFNN models are much easier to be designed and trained than other neural networks. Furthermore, RBFNN models are less sensitive to noise and faster than other neural networks.

RBFNN is one of neural network models that has been used in SDEE. Lopez-Martin [20] identified ten studies that used RBFNN in SDEE. The main limitations of the RBFNN software effort estimation models identified by Lopez-Martin [20] are:

- The majority of these models use MMRE as an evaluation criterion, which is considered biased and not recommended as stated by Shepperd and MacDonell [23].
- Most of these models did not use statistical tests to validate their models and this process is very important [30].
- Most of these models did not report the parameters of the RBFNN, and some reported but without any justification.

## 2.3 General Regression Neural Network (GRNN)

The GRNN is a type of neural network that was proposed by Specht [31]. A GRNN network applies regression on continuous output variables. An GRNN is composed of four layers as depicted in Figure 3. The first layer represents the input layer in which each predictor (aka independent variable) has a neuron. The second layer is fed from the input neurons.

The second layer is composed of pattern neurons such that each neuron represents a training row. Each neuron applies the RBF function using the sigma "$\sigma$" to compute the Euclidean distance from the input vector *X* to the neuron's center. The output of the second layer feeds the third layer (summation neurons).

The third layer is composed of two neurons. The first one is called the denominator summation, which adds the values of the weights coming from each of the second-layer neurons. The other neuron is the numerator summation that adds the weights multiplied by the actual output value of each pattern neurons.

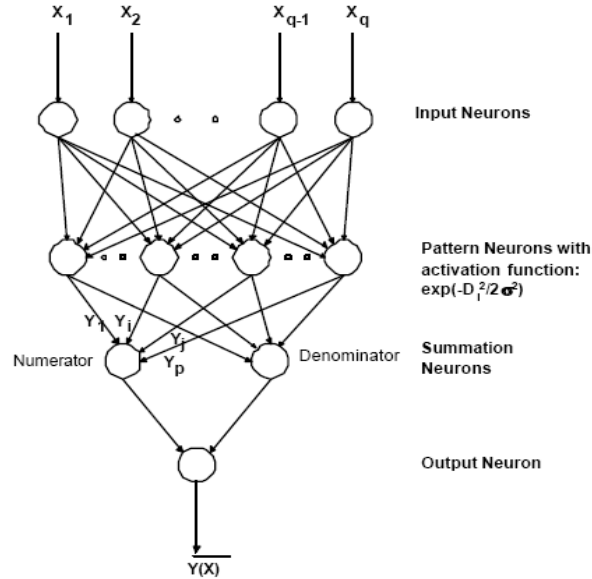

Figure 3: General Regression Neural Network [7]

The fourth layer contains the output neuron, which is the predicted output value. This is calculated based on the division of the value stored in the numerator neuron by the value stored in the denominator neuron.

The application of GRNN models in SDEE is limited. Lopez-Martin applied GRNN models in two studies: [19] and [32]. In the first study, short-scaled projects are used, whereas ISBSG industrial projects were used in the second one. Another study was reported by Nassif et al. [33] where projects based on use case points (UCP) were used. The evaluation of the GRNN models was based on their comparison with multiple linear regression models based on the Mean Magnitude of Relative Error (MMRE) or the Mean Magnitude of Error Relative to the estimate (MMER).

## *2.4 Cascade Correlation Neural Network (CCNN)*

The CCNN was proposed by Fahlman et al. in 1990 [34]. A CCNN network, which is also known as a self-organizing networks, is composed of an input, hidden, and output layers. When the training process starts, a CCNN network is only composed of an input and output layers as depicted in Figure 4. Each input is connected to each output. The bias is represented by the "+1". The "x" sign represents a modifiable weight between the input and the output neurons. In the second stage, neurons are added to the hidden layer one by one. Figure 5 shows a CCNN network with one neuron. The existing input neurons and the output of the previous added neuron feeds the inputs of a new hidden neuron as shown in Figure 5. The square sign denotes fixed weights. When a new hidden neuron is added, the training algorithm tries to reduce the residual error by maximizing the magnitude of the correlation between the new hidden neuron's output and the residual error. If the residual error is not reduced by adding a new hidden neuron, the training process will stop.

Only one study conducted by Nassif at al. [21] is reported where a CCNN model is used in SDEE. In this study, projects based on use case points were used. The CCNN model was compared to a multiple linear regression model using the MMER criterion.

As seen above, studies that used MLP models claimed that MLP models are the best among neural network models. This claim is applied to those who used RBFNN, GRNN, and CCNN models as well. There are several reasons behind this discrepancy in results. These include:

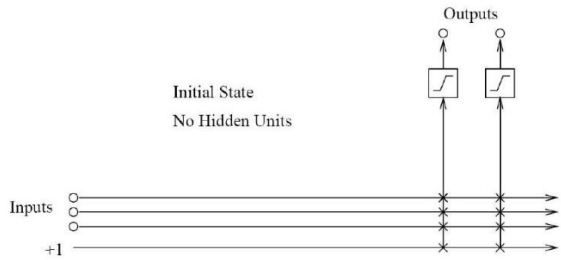
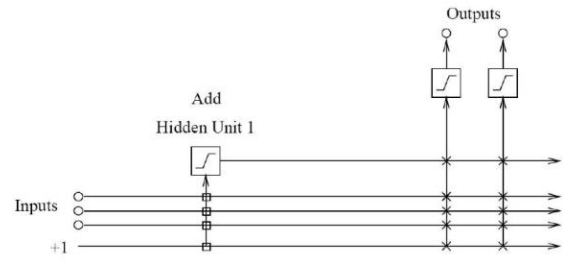

Figure 4: CCNN with no Hidden Neurons [34]         Figure 5: CCNN with One Hidden Neuron [34]

1. Choosing datasets in model evaluation: This is a very important issue in SDEE. All models are evaluated using datasets and the performance of such models largely depends on the characteristics of the dataset being used. For instance, the performance of a model might excel on one dataset and be detrimental on another dataset. For this reason, SDEE models should be evaluated based on more than one dataset.

2. Splitting the dataset into training vs. testing data points: Another issue in evaluating SDEE models is how to split the dataset into training and testing datasets. The traditional method, which most of the above models use, is to randomly split the dataset into 70% for training and 30% for testing. This method, despite its fame, is not recommended for two main reasons. First, by keeping random changing of the 70% / 30% split of a dataset, different results will be obtained. This will lead researchers to choose a split that yields better results. Another reason for not recommending the 70% / 30% split method is that it is not replicable. As an alternative to this split method, Leave-One-Out (LOO) and cross-validation techniques are recommended.

3. Choosing performance evaluation criteria: There are more than ten evaluation performance criteria that are being used in the evaluation of SDEE models. Some of them are considered biased and should not be used. For instance, the MMRE criterion is in favour of SDEE models that tend to underestimate (predicted effort is less than actual effort). On the contrary, the MMER criterion is in favour to models that overestimate. In spite of its criticism, around 80% of SDEE models have used biased criteria such as MMRE.

In this research, we conduct an unbiased comparative study to compare MLP, RBFNN, GRNN, and CCNN models. First, we used five industrial datasets from the ISBSG. The next section explains the algorithm used to extract and filter these datasets. Second, we used the MAR as an unbiased evaluation performance criteria [23] to compare the four models. So the four models were trained on the same datasets using the cross-validation method, then evaluated using new datasets that were not included in the training stage based on the MAR criterion. Further, we use statistical tests to check if a model is statistically different than other models.

## 3. Datasets

To perform unbiased comparison between neural network models, we used industrial datasets published by the ISBSG Release 11 [22]. ISBSG Release 11 datasets contain more than 5,000 cross-company projects from all over the world. The main characteristics / issues of ISBSG datasets are as follows:
1. ISBSG projects were developed using different platforms, different programming languages, and different software development life cycle models.
2. Different metrics for software size are used. These include SLOC, IFPUG, COSMIC, etc.
3. Each project is ranked as "A", "B", or "C" based on the quality of this project. The ISBSG guideline recommends dismissing projects whose rank is different from "A" and "B".
4. Many rows (projects) contain missing data.
5. There are more than 100 columns (features), some of them (such as project number, project date, etc.) are not correlated to the output (software effort). Some of the features are correlated but are statistically insignificant as determined by statistical tests.
6. There are two main types of development: new development and enhancement.

7. The dataset is very heterogeneous and the productivity (ratio between software effort and software size) varies dramatically even with the same size metric. For instance, for projects of metric size IFPUG, the productivity varies between 0.1 and 621. For example, if a project size is 100 units, the effort required to develop this project varies between 10 hours (if productivity is 0.1) and 6,210 hours (if productivity is 621). This is a main concern and this issue needs to be addressed.

Based on the above characteristics of ISBSG, we developed an algorithm that is replicable to filter the ISBSG Release 11 and generate five subsets (datasets). We used the guidelines provided by the ISBSG to generate five subsets from the main population. First, regarding the features, we chose IFPUG Adjusted Function Points (AFP), development type of "new development", development platform, language type, resource level and normalised work effort. The latter (normalised work effort) is the output of the model where the other features are the inputs. It is worth mentioning that among the inputs of the model, only the software size (AFP) is a continuous variable where the others are categorical variables. As recommended by ISBSG, only projects of quality rated "A" or "B" with no missing data were considered.

To solve the issue discussed above in issue 7, five different subsets were taken based on the value of the productivity. The first subset is chosen when the values of productivity vary between 0 and 4.9 inclusive. The second one is when the productivity values are between 5 and 9.9, and so on. After five datasets are generated, each dataset is divided into training datasets and testing dataset. The methodology used for datasets splitting is to sort the projects based on chronological order; the oldest 70% projects were used for training and the newest 30% were used for testing. This method resembles the situation in real life where historical projects are used for training to predict the effort of a new project. Please note that this method of splitting is different from the random 70% / 30% discussed above since the splitting in this method is not random and this method is replicable.

The algorithm used to generate five subsets (datasets) from the main one is displayed below:

**Algorithm for Development Type = New Development:**
1. Total number of projects (data points) = 5,052
2. Consider only Data Quality A and B (column B in the dataset) ➔ 4,474 projects left
3. Consider only Count Approach = IFPUG (column E in the dataset) ➔ 3,614 projects left
4. Select attributes Adjusted Function Points (column G), Normalised Work Effort (Column J), Development Type (column AF), Development Platform (column BQ), Language Type (column BR), and Resource Level (Column CU). Please note that Normalised Work Effort is the output of the model (Dependent Variable) where the other attributes are the input of the model (independent variables). Please also note that Development Type will not be an input because all projects have the same development type ➔ 3,614 projects left
5. Delete all rows that contain missing data related to each of the above features (part 4) ➔ 2,815 projects left
6. Select Development Type = New Development ➔ 951 projects left
7. Divide the main dataset into five main datasets based on productivity value:
    a. Dataset1: where productivity is between 0 and 4.9 ➔ 288 projects
    b. Dataset2: where productivity is between 5 and 9.9 ➔ 260 projects
    c. Dataset3: where productivity is between 10 and 14.9 ➔ 138 projects
    d. Dataset4: where productivity is between 15 and 19.9 ➔ 101 projects
    e. Dataset5: where productivity is greater than 20 ➔ 164 projects
8. Sort each dataset based on chronological order (based on the attribute 'Year of Project' from oldest to newest). Split the dataset into Training Dataset (the oldest 70% of the dataset) and Testing Dataset (the recent 30% of the dataset)
    a. Dataset1: 202 training projects (from 1990 to 2001) and 86 testing projects (from 2001 to 2006)
    b. Dataset2: 182 training projects (from 1994 to 2002) and 78 testing projects (from 2002 to 2008)
    c. Dataset3: 97 training projects (from 1989 to 2002) and 41 testing projects (from 2002 to 2007)
    d. Dataset4: 71 training projects (from 1998 to 2003) and 30 testing projects (from 2004 to 2006)
    e. Dataset5: 115 training projects (from 1993 to 2000) and 49 testing projects (from 2000 to 2008)

Figures 6 to 10 show the relationship between software size (Adjusted Function Points) and software effort (Person-Hours) in each of the five datasets.

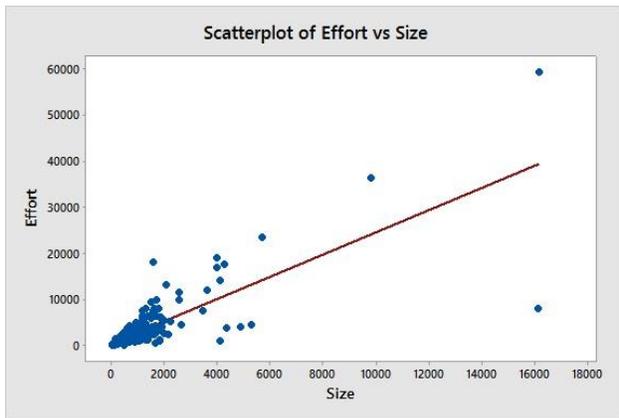
Figure 6: Dataset1

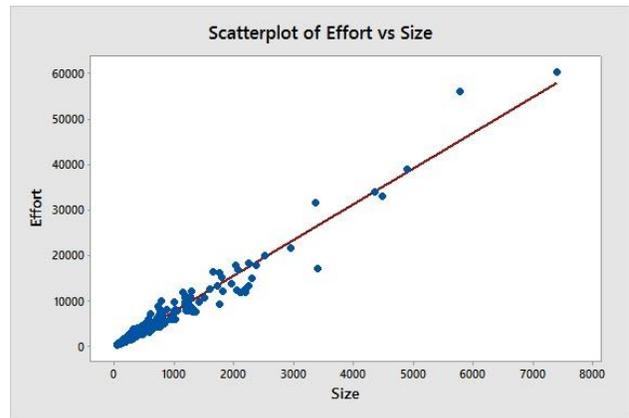
Figure 7: Dataset2

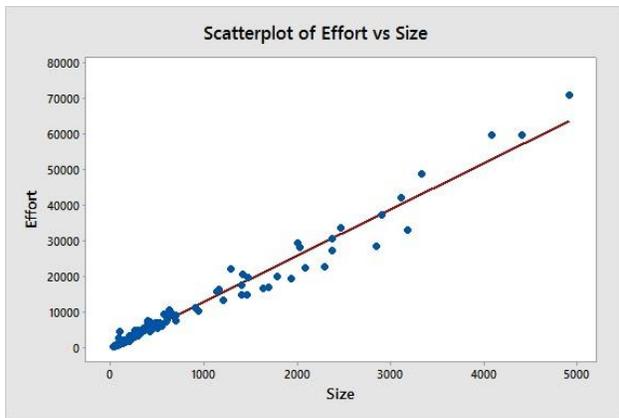
Figure 8: Dataset3

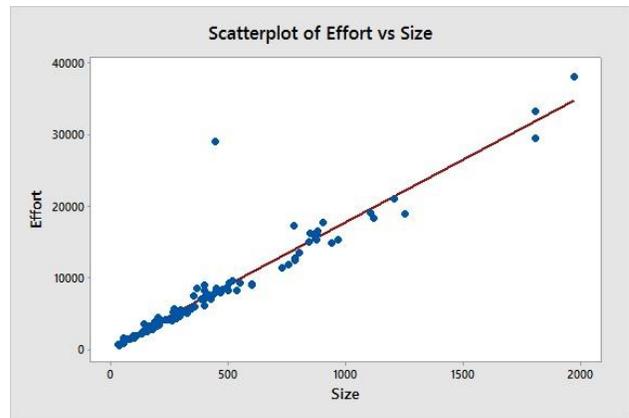
Figure 9: Dataset4

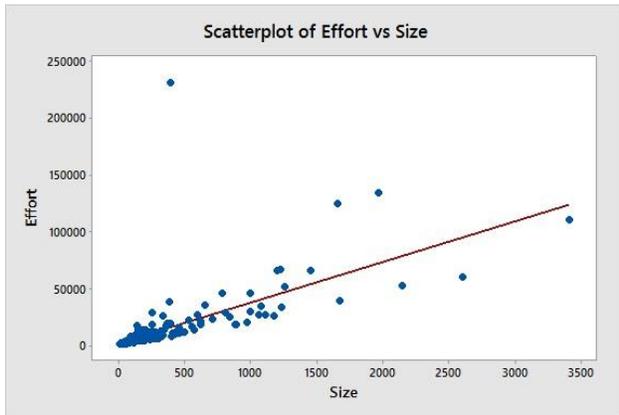
Figure 10: Dataset5

Table 1 shows the descriptive statistics for the output "Effort" for each dataset.

Based on Table 1, we notice that the datasets 1 and 5 are not normally distributed based on the values of the Skewness and Kurtosis. However, the distribution in datasets 3 and 4 is more normal than other datasets. We also conclude from Figures 6 to 10 that datasets 1, 4 and 5 contain some outliers.

Table 1: Descriptive Statistics for Variable Effort (Person-Hour)

|          | Count | Mean   | StDev  | Minimum | Median | Maximum | Skewness | Kurtosis |
|----------|-------|--------|--------|---------|--------|---------|----------|----------|
| Dataset1 | 288   | 2,699  | 5,055  | 26      | 1,154  | 59,262  | 6.71     | 62.19    |
| Dataset2 | 260   | 5,209  | 7,297  | 254     | 2,761  | 60,270  | 4.2      | 23.77    |
| Dataset3 | 138   | 8,556  | 12,448 | 366     | 3,193  | 71,118  | 2.72     | 8.3      |
| Dataset4 | 101   | 8,064  | 7,100  | 498     | 5,565  | 38,095  | 1.89     | 4.28     |
| Dataset5 | 164   | 15,227 | 26,158 | 686     | 7,579  | 230,514 | 5        | 32.39    |

## 4. Evaluation Performance Criteria

Shepperd and MacDonell [23] criticized the use of the MMRE and MMER criteria despite their popularity because these criteria are biased; the MMRE criterion favours the models that underestimate whereas the MMER criterion favours the models that overestimate. For this reason, we have been using the unbiased criterion Mean Absolute Residual (MAR) to compare the accuracy of neural network models. MAR is expressed as:

$$MAR = \frac{|Ea - Ep|}{n} \quad (3)$$

Where $Ea$ is the actual effort, $Ep$ is the predicted effort and $n$ is the number of observations. Equation (3) will be used while answering RQ1.

To answer RQ2 (which model tends to overestimate or underestimate), we used the mean of the residuals (MR) criterion which is expressed in Equation (4):

$$MR = \frac{Ea - Ep}{n} \quad (4)$$

A negative MR indicates that such model tends to overestimate. On the other hand, a positive value for MR indicates that the model is underestimating.

The question RQ3 will be addressed in Section 6.

## 5. Models Training and Testing

As discussed in Section 3, we are using five datasets to compare the four models (MLP, RBFNN, GRNN, and CCNN). Each dataset is split into a training dataset and a testing dataset. All models are trained based on the same datasets using the 10-fold cross validation method. Each model has four inputs, which are: (1) software size (AFP), (2) development platform, (3) language type, and (4) resource level. The output of the model is software effort. After the models have been trained, they were tested on the same datasets that were not included in the training stage.

For the MLP model, we used the conjugate gradient algorithm [27]. One of the most important steps in training an MLP model is to determine the number of the neurons in the hidden layer. In the training process of the MLP model, the initial number of the neurons in the hidden layer was set to one, and then it was incrementally increased by one until optimal results were achieved. If the training error decreases and the validation error begins to increase, the training was stopped to avoid overfitting. At each number of hidden neurons, the residual variance was calculated. The residual variance determines how well the model fits the dataset. The smaller the variance, the more accurate the model is. The parameters of the model are: Convergence Tolerance = $1.0e^{-5}$, Maximum Iterations = 10,000, Minimum Gradient = $1.0e^{-6}$, and Minimum Improvement Delta = $1.0e^{-6}$. The number of hidden neurons in Dataset1, Dataset2, Dataset3, Dataset4, and Dataset5 are 11, 4, 4, 2, and 9, respectively.

For the RBFNN model, the number of hidden neurons also varied based on the dataset. For each neuron, each input variable had a width (spread) and a centre. The RBFNN model was trained based on the algorithm proposed by Chen et al. [35]. The number of the neurons in the hidden network starts by one and it will be incrementally increased by one until it achieves the optimal training results. Optimal results are achieved when the average error in the training stage and the validation error is minimal. The number of hidden neurons in Dataset1, Dataset2, Dataset3, Dataset4, and Dataset5 are 10, 5, 5, 5, and 8, respectively.

For the GRNN, the most important parameter is the spread value. If the spread value is very small, the training error will be small but the validation error will be high and this leads to overfitting. When the spread value increases, the training error increases where the validation error decreases to a point where both the training and validation errors become equal. In GRNN models, a model can have a spread value or each input variable can have a spread value and this is the case in our model. The spread values of software size (AFP) in dataset1, dataset2, dataset3, dataset4, and dataset5 are 0.95, 0.48, 0.88, 0.26, and 1.62, respectively.

For the CCNN, the number of the hidden neurons starts at zero, then it is increased until optimal results are achieved. Optimal results are achieved when the validation residual is minimal. In CCNN models, it is possible that the number of hidden neurons is zero. The number of hidden neurons in dataset1, dataset2, dataset3, dataset4, and dataset5 are 6, 0, 0, 0, and 0, respectively.

When the neural network models were trained, they were tested on a new dataset, as described in Section 3. The values of MAR and MR of each model in each dataset are reported in Table 2. Figures 11 to 15 show the interval plot of the MAR for each dataset. Table 3 displays the significance (number 1 indicates the most significant) of each input variable based on each model for each dataset. In order for the values to fit in the table, the inputs software size, development platform, language type, and resource level are labeled as AFP, DP, LT, and RL, respectively. Moreover, MLP, GRNN, RBFNN, and CCNN are labeled as M, G, R, and C, respectively.

Table 2: Values of MAR and MR

|  | MAR | | | | MR | | | |
| --- | --- | --- | --- | --- | --- | --- | --- | --- |
|  | MLP | GRNN | RBFNN | CCNN | MLP | GRNN | RBFNN | CCNN |
| Dataset1 | 1527 | 1265 | 1124 | 1479 | -133 | -174 | -172 | -290 |
| Dataset2 | 667 | 1042 | 569 | 544 | 297 | 78 | 197 | 368 |
| Dataset3 | 1039 | 1871 | 588 | 760 | -138 | -384 | 20 | 69 |
| Dataset4 | 1321 | 1373 | 1301 | 1066 | -621 | -1086 | -938 | -771 |
| Dataset5 | 7185 | 7087 | 8283 | 6833 | -1552 | -1925 | -3377 | -1831 |

Table 3: Significance of Variables

|  | AFP | | | | DP | | | | LT | | | | RL | | | |
| --- | --- | --- | --- | --- | --- | --- | --- | --- | --- | --- | --- | --- | --- | --- | --- | --- |
|  | M | G | R | C | M | G | R | C | M | G | R | C | M | G | R | C |
| Set1 | 1 | 1 | 1 | 1 | 3 | 4 | 4 | 3 | 2 | 2 | 2 | 2 | 4 | 3 | 3 | 4 |
| Set2 | 1 | 1 | 1 | 1 | 2 | 2 | 2 | 2 | 3 | 3 | 3 | 4 | 4 | 4 | 4 | 3 |
| Set3 | 1 | 1 | 1 | 1 | 3 | 2 | 2 | 2 | 2 | 3 | 3 | 4 | 4 | 4 | 4 | 3 |
| Set4 | 1 | 1 | 1 | 1 | 2 | 2 | 3 | 2 | 4 | 4 | 3 | 4 | 3 | 3 | 4 | 3 |
| Set5 | 1 | 1 | 1 | 1 | 2 | 2 | 2 | 2 | 4 | 4 | 4 | 4 | 3 | 3 | 3 | 3 |

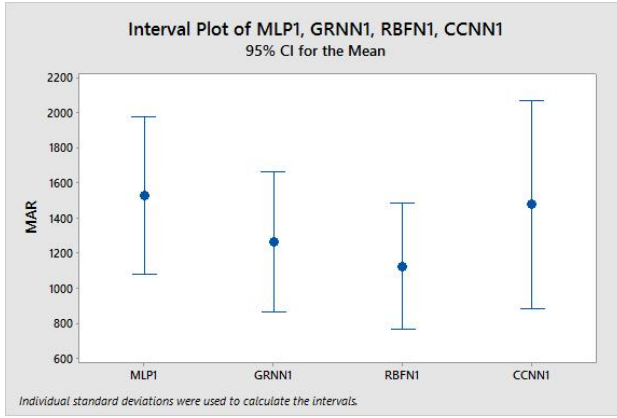

Figure 11: MAR Interval plot - Dataset1

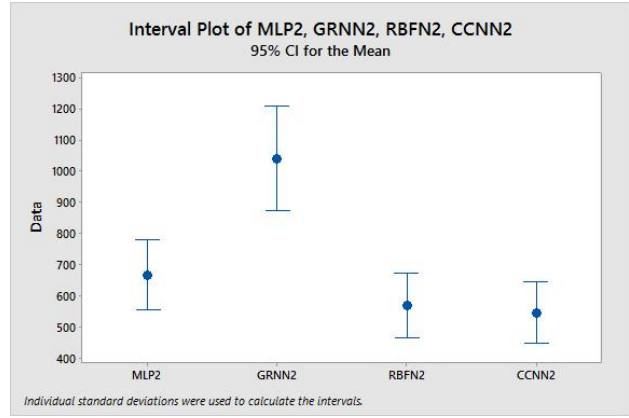

Figure 12: MAR Interval plot - Dataset2

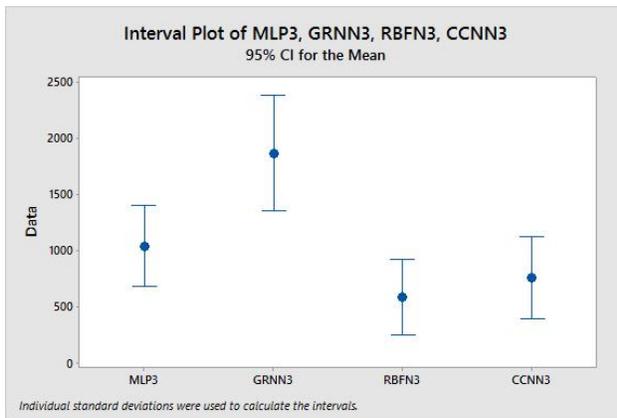

Figure 13: MAR Interval plot - Dataset3

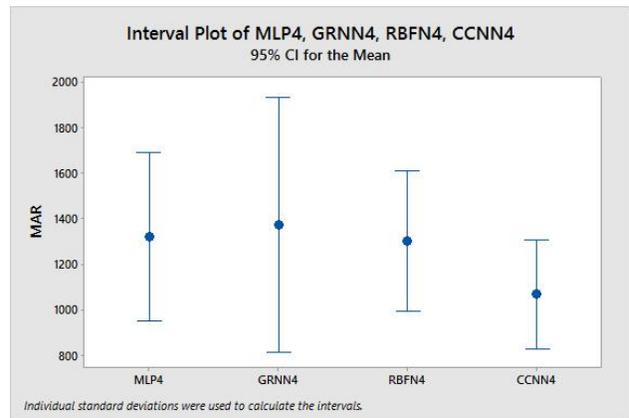

Figure 14: MAR Interval plot - Dataset4

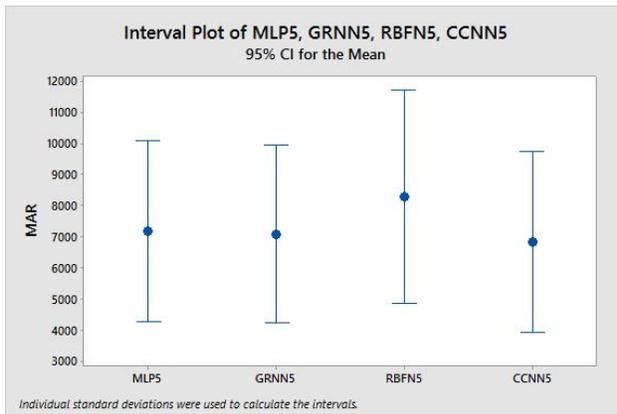

Figure 15: MAR Interval plot - Dataset5

## 6. Discussion

In this section, we will discuss the results depicted in Section 5 and try to answer the research questions listed in Section 1. Regarding the values of the MAR criterion in Table 2, it is clear (shaded cells) that the CCNN model and the RBFNN model outperform the MLP and the GRNN models since they have the lowest MAR values. Remarkably, the CCNN model excelled in three datasets, followed by the RBFNN which was superior in two datasets. These results are very interesting because the attractiveness of applying CCNN in SDEE is much lower than the attraction of other neural network models such as the MLP and RBFNN models. This indicates that more attention should be given to the CCNN in SDEE. However, this observation should not be deemed credible unless supported with statistical tests.

To enrich the investigation of the MAR results of the CCNN model, we conducted statistical tests to see if the CCNN model is statistically different from other models in the five datasets. We also applied statistical tests to see if the RBFNN, the second winning model is statistically different from other models. We applied a normality test on the MAR and found that data were not normally distributed. For this reason, we applied the Wilcoxon test, which is a non-parametric test; this type of tests should be applied when the distribution is not normal. Table 4 shows the results of the Wilcoxon test based on the 95% Confidence Interval. Shaded cells indicate that the model is statistically different at 95% CI.

Table 4: Wilcoxon Test for the CCNN and RBFNN Models

| Dataset | P value at 95% Confidence Interval (CI) | | | | |
| --- | --- | --- | --- | --- | --- |
| | CCNN vs. MLP | CCNN vs. GRNN | CCNN vs. RBFNN | RBFNN vs. MLP | RBFNN vs. GRNN |
| Dataset1 | 0.33 | 0.22 | 0.43 | 0.07 | 0.10 |
| Dataset2 | 0.10 | 0.00 | 0.79 | 0.17 | 0.00 |
| Dataset3 | 0.00 | 0.00 | 0.18 | 0.00 | 0.00 |
| Dataset4 | 0.55 | 0.70 | 0.30 | 0.56 | 0.19 |
| Dataset5 | 0.88 | 0.50 | 0.39 | 0.38 | 0.96 |

The null and the alternative hypothesis of the Wilcoxon test of the CCNN model are as follows:

$H_0$: The CCNN mode is not statistically different from the other models at 95% CI.

$H_1$: The CCNN mode is statistically different from the other models at 95% CI.

The null and the alternative hypothesis of the Wilcoxon test of the RBFNN model are as follows:

$H_0$: The RBFNN mode is not statistically different from the other models at 95% CI.

$H_1$: The RBFNN mode is statistically different from the other models at 95% CI.

If the p-value is less than 0.05, then we reject the null hypothesis. Otherwise, we fail to reject the null hypothesis. Based on Table 4, we notice that The CCNN model is not statistically different from the RBFNN model in all datasets. Moreover, both models fail to be statistically different from other models in all datasets. Based on Table 2, we notice that the CCNN model slightly outperforms the RBFNN model based on the MAR criterion but based on statistical tests, these models are almost identical.

### 6.1 Answers to Research Questions

RQ1: Which of the above neural network models (MLP, GRNN, RBFNN, and CRNN) has the lowest MAR?

Based on the results of the experiments, we conclude that the CCNN model outperforms other models in 60% of the datasets when the MAR criterion was used. The RBFNN is ranked second since it outperformed other models based on 40% of the datasets. However, when we applied the Wilcoxon non-parametric test, we noticed that the CCNN model was not statistically different from other models in all datasets.

RQ2: Which neural network model tends to overestimate and which tends to underestimate?

Based on the MR values in Table 2, we conclude that the MLP and GRNN models tend to overestimate based on 80% of the datasets, followed by the RBFNN and CCNN models which tend to overestimate based on 60% of the datasets. Moreover, we notice that the peak of overestimation of the effort occurs when Dataset5 is used. Figure 10 shows that Dataset5 contains an outlier and this may affect the performance of the neural network models. Some models are more sensitive to outliers than other models. Furthermore, Datset5 is not normally distributed, as opposed to Dataset3; in the latter we notice that there are no outliers and the distribution is more normal than other datasets. The RBFNN model looks as if it is more sensitive to outliers than other models and when data are not normally distributed; this happens because it is the highest among other models that tend to overestimate. On the other hand, the RBFNN model is not overestimating nor underestimating in Dataset3 where there are no outliers and data are more normally distributed.

RQ3: Does the significance of model inputs vary from model to model?

In Table 3, we notice that software size is the most significant attribute in all models based on the five datasets. However, the significance of other features (development platform, language type, and resource type) varies from model to model even within the same dataset. This outcome contradicts with non-machine learning models, such as statistical model, when the stepwise regression, for example, can be used to determine the significance of each independent variable (features) based on a certain level of significance (e.g., $\alpha = 0.05$). In other words, the stepwise regression test should not be used with neural network models to determine the significance of the model inputs.

## 7. Threats to Validity

In this section we mention some threats that might have affected the validity of the neural network models, which indirectly have an impact on the comparisons. These include:

1. We did not use the MMRE as a performance evaluation criterion as suggested by Shepperd and MacDonell [23] because it is biased and favours models that underestimate. However, if we had used MMRE, this would have helped in comparing models developed in other studies since the MMRE is the most used criterion in similar research.

2. We used datasets from the ISBSG to conduct the experiments. These datasets came from different companies. This process is called cross-company as opposed to within-company where projects are developed in the same company. There is a debate as to whether using cross-company projects is significantly different form using within-company projects [36]. Nonetheless, not using within-company projects in this research can be considered a threat.

3. We used the 10-fold cross validation technique in the training process of the models. Some studies prefer using the Leave-One-Out (LOO) method, especially with small datasets.

## 8. Conclusions and Future Work

This research has been carried out to compare four neural network models: (1) Multilayer Perceptron (MLP), (2) General Regression Neural Network (GRNN), (3) Radial Basis Function Neural Network (RBFNN), and (4) Cascade Correlation Neural Network (CCNN). We used five datasets extracted from the ISBSG. The performance criterion used was the Mean Absolute Residual (MAR). Each model has four inputs: (1) software size, (2) development platform, (3) language type, and (4) resource level. Software effort was the output of the model. The comparison was based on three research questions:

RQ1: Which of the above neural network models (MLP, GRNN, RBFNN, and CRNN) has the lowest MAR?

RQ2: Which neural network model tends to overestimate and which tends to underestimate?

RQ3: Does the significance of model inputs vary from model to model?

Based on the resulting discussions in Sections 5 and 6, we conclude the following:

1. The CCNN model outperformed other models based on 60%, of the datasets followed by the RBFNN model that surpassed other models based on 40% of the datasets.

2. We conducted statistical tests and found that the CCNN model, despite its high performance, is not statistically different from other models.

3. The four models tend to overestimate the effort required to develop software projects.

4. The RBFNN is sensitive to outliers and when data are not normally distributed.

5. Software size is the most significant attribute in all models. The significance of other attributes varies from model to model.

Future work will focus on conducting the comparison of models using within-company projects. LOO validation technique and other performance evaluation criteria can also be considered in the future.

## Compliance with Ethical Standards

Ali Bou Nassif would you to thank the University of Sharjah for supporting this research.

Luiz Fernando Capretz and Danny Ho would like to thank the Natural Sciences and Engineering Research Council of Canada (NSERC) for their support of this work through a Discovery Grant -Team.

Conflict of Interest: The authors declare that they have no conflict of interest.

Informed consent: This study does not involve any human participants and/or animals.